\documentclass{optica-article}

\journal{oe}

\articletype{Research Article}

\usepackage{lineno}
\linenumbers

\usepackage{graphicx}
\usepackage{xcolor}
\usepackage{cite}
\usepackage{subcaption}
\usepackage{amsmath}
\usepackage{xspace}
\usepackage{pgfplots}
\pgfplotsset{width=10cm,compat=1.9}
\usepgfplotslibrary{colorbrewer}

\definecolor{PlotRed}{HTML}{e41a1c}
\definecolor{PlotBlue}{HTML}{377eb8}
\definecolor{PlotGreen}{HTML}{4daf4a}
\definecolor{PlotPurple}{HTML}{984ea3}
\definecolor{PlotOrange}{HTML}{ff7f00}
\definecolor{gridGrey}{HTML}{dddddd}

\pgfdeclarelayer{background layer}
\pgfdeclarelayer{foreground layer}
\pgfsetlayers{background layer,main,foreground layer}

\makeatletter
\DeclareRobustCommand\onedot{\futurelet\@let@token\@onedot}
\def\@onedot{\ifx\@let@token.\else.\null\fi\xspace}

\makeatother

\begin{document}
\nolinenumbers

\title{Equalization in Dispersion-Managed Systems Using Learned Digital Back-Propagation}

\author{Mohannad Abu-romoh,\authormark{1,*} 
Nelson Costa,\authormark{2} 
Yves Jaou\"en,\authormark{1} 
Antonio Napoli,\authormark{3} 
Jo\~ao Pedro,\authormark{2,4} 
Bernhard Spinnler,\authormark{3} 
and Mansoor Yousefi\authormark{1}}

\address{\authormark{1}Institut Polytechnique de Paris, T\'el\'ecom Paris, 19 Place Marguerite Perey, 91120 Palaiseau, France \\
\authormark{2}Infinera Unipessoal Lda, 2790-078 Carnaxide, Portugal \\
\authormark{3}Infinera, Munich, Germany \\
\authormark{4}Instituto de Telecomunicações, Instituto Superior Técnico, 1049-001 Lisboa, Portugal}

\email{\authormark{*}mohannad.aburomoh@telecom-paris.fr} 

\begin{abstract}
In this paper, we investigate the use of the learned digital back-propagation (LDBP) for equalizing dual-polarization fiber-optic transmission in dispersion-managed (DM) links. LDBP is a deep neural network that optimizes the parameters of DBP using the stochastic gradient descent. We evaluate DBP and LDBP in a simulated WDM dual-polarization fiber transmission system operating at the bitrate of 256 Gbit/s per channel, with a dispersion map designed for a 2016 km link with 15\% residual dispersion.
Our results show that in single-channel transmission, LDBP achieves an effective signal-to-noise ratio improvement of 6.3 dB and 2.5 dB, respectively, over linear equalization and DBP. In WDM transmission, the corresponding $Q$-factor gains are 1.1 dB and 0.4 dB, respectively. Additionally, we conduct a complexity analysis, which reveals that a frequency-domain implementation of LDBP and DBP is more favorable in terms of complexity than the time-domain implementation.
These findings demonstrate the effectiveness of LDBP in mitigating the nonlinear effects in DM fiber-optic transmission systems.
\end{abstract}


\section{Introduction}

Before the advent of the coherent detection in optical fiber communication systems, chromatic dispersion (CD) was compensated in the optical domain through in-line dispersion management. In such systems, an optical link typically comprises a cascade of a single-mode fiber (SMF), followed by a dual-stage Erbium-doped fiber amplifier (EDFA) with a mid-stage dispersion-compensating fiber (DCF). In this scenario, the residual CD in the link forms a dispersion map that requires careful optimization. These systems have been utilized for transmitting data rates typically around 10 Gb/s per wavelength, with a few instances of upgrades to 40 Gb/s\cite{Napoli:2014}.

With the introduction of the coherent receiver (RX), the accumulated CD can now be fully compensated in the electrical domain using digital signal processing (DSP). As a result, dispersion management is no longer a necessity, leading to the evolution of optical links to non-dispersion-managed (NDM) systems. However, commercial dispersion-managed (DM) links continue to be in operation today, since upgrading deployed solutions to state-of-the-art systems can be particularly costly, especially in submarine fiber links. Furthermore, advances in nonlinear mitigation in recent decades may address some of the limitations of the DM systems.

Coherent systems depend on DSP at the receiver (RX) to equalize transmission impairments in the electrical domain \cite{DSP_savory}. While these systems effectively compensate for linear channel impairments, nonlinearity mitigation remains challenging. The pursuit of the nonlinearity compensation (NLC) and associated DSP algorithms has led to the proposal of several equalization techniques to mitigate nonlinear impairments \cite{Clara_lit_review,Sidelnikov:18}, including Volterra series-based equalization \cite{Gao-Volterra} and digital back-propagation (DBP) \cite{Ezra_Ip_DBP,millar_DBP,Mateo:10,Czegledi:17}.

DBP compensates for deterministic nonlinear effects by digitally simulating propagation at the receiver (RX) using negated fiber parameters \cite{Ezra_Ip_DBP}. However, the effectiveness of DBP is hindered by both hardware limitations, which impose restrictions on the complexity of DBP, and signal interactions between adjacent channels. Thus, advanced equalization techniques are essential for improving system performance and reducing complexity.

Research has focused on optimizing DBP algorithms for lower complexity or better performance in various transmission scenarios \cite{pro_dig_bac,Ove_per_lim}. For instance, channel parameters used in DBP were numerically optimized for DM and NDM systems in \cite{Com_tra_imp}. Combining transmitter-side DBP with frequency referenced carriers has been shown to double the reach of the transmission, as demonstrated in \cite{Temprana:15}. Optimal DBP step sizes for polarization division multiplexed transmission systems were studied in \cite{DBP_opt_sts}. Furthermore, several DBP variants have been proposed to reduce complexity or enhance performance, such as correlated-DBP \cite{cor_dbp_bas}, which accounts for the correlation between neighboring signal samples. Dispersion folded-DBP \cite{Zhu:11,4907020,Zhu:12,Eff_fib_non} was first proposed for zero-residual dispersion but later extended to any dispersion map. Filtered DBP \cite{Du:10,Rafique:11} introduces a parameterized low-pass filter (LPF) in the nonlinear step to improve phase tracking. Enhanced DBP \cite{6964122,647ecab0e} is an extension of FDBP that considers interactions between channel signals and adjacent channels. Coupled-Channel Enhanced DBP \cite{9489431} achieves optimal cross-phase modulation (XPM) equalization in WDM systems.
Recently, neural networks (NNs) have been employed to enhance DBP. Learned-DBP (LDBP) \cite{Hager_butler_LDBP,Hagers_phys_based,Opt_express_LDBP,Hagers_first,Abu-romoh:22} uses a deep NN inspired by the split-step Fourier method (SSFM) computational graph, optimized using stochastic gradient descent (SGD). LDBP treats the LPF filter taps in DBP as free parameters optimized by the NN.
For NDM systems, LDBP was extended to dual-polarization transmission \cite{Hager_butler_LDBP} and experimentally demonstrated with one layer per span (LpS) \cite{Hagers_phys_based}. Generalized DBP (GDBP) \cite{APTL_NN} employs a deep NN to parameterize DBP, specializing the system by combining NN training with adaptive DSP.

The main objective of this paper is to investigate the potential of repurposing legacy DM links for use in coherent transmission systems. We do not claim that replacing NDM with DM systems is superior; rather, we aim to explore the possibility of adapting the already-deployed DM systems that are typically used for IM/DD to enable coherent transmission. To achieve this goal, we use machine learning tools to optimize DBP, specifically the Learned-DBP method. We start by developing a DBP variant that is suitable for DM systems with a fractional number of steps per span (StpS), which can be applied to DM links with arbitrary dispersion maps. The resulting DBP is then used as the blueprint for LDBP, and we evaluate the performance of both DBP and LDBP in a realistic long-haul dual-polarization WDM DM transmission system with $M$-ary quadrature amplitude modulation ($M$-QAM), considering various channel effects such as loss, CD, Kerr nonlinearity, polarization-mode dispersion (PMD), amplified spontaneous emission (ASE) noise, and laser phase noise (PN).

In single-channel transmission, LDBP improves the effective signal-to-noise ratio (SNR$_{\text{eff}}$) by 6.3 dB and 2.5 dB compared to linear equalization (LE) and conventional DBP, respectively. In WDM transmission, LDBP enhances the Q-factor by 1.1 dB and 0.4 dB compared to LE and DBP, respectively. We present both time- and frequency-domain implementations of LDBP (TD- and FD-LDBP) and examine the impact of fiber parameter variations due to aging, as well as laser PN, on LDBP performance. Our analysis shows that PN results in a Q-factor penalty of less than 0.2 dB when the laser linewidth is below 200 kHz. Furthermore, we demonstrate that LDBP's performance gains over DBP are approximately maintained even when the model is retrained with updated fiber parameters post-aging. This work extends the findings in \cite{Abu-romoh:22} by providing further valuable insights into improving data rates in DM systems utilizing coherent detection and LDBP.

The paper is structured as follows: In Section II, we describe the optical fiber signal propagation model. In Section III, we present the DM-adapted DBP, while Section IV introduces its learned version. In Section V, we compare the performance of LE, DBP, and LDBP in four different setups and provide a complexity analysis in Section VI. Finally, Section VII offers concluding remarks.

\section{Dual-Polarization Optical Fiber System Model}

\begin{figure}[t]
    \centering
    \includegraphics[width=1\linewidth,trim={0.75cm 0 1.2cm 0},clip]{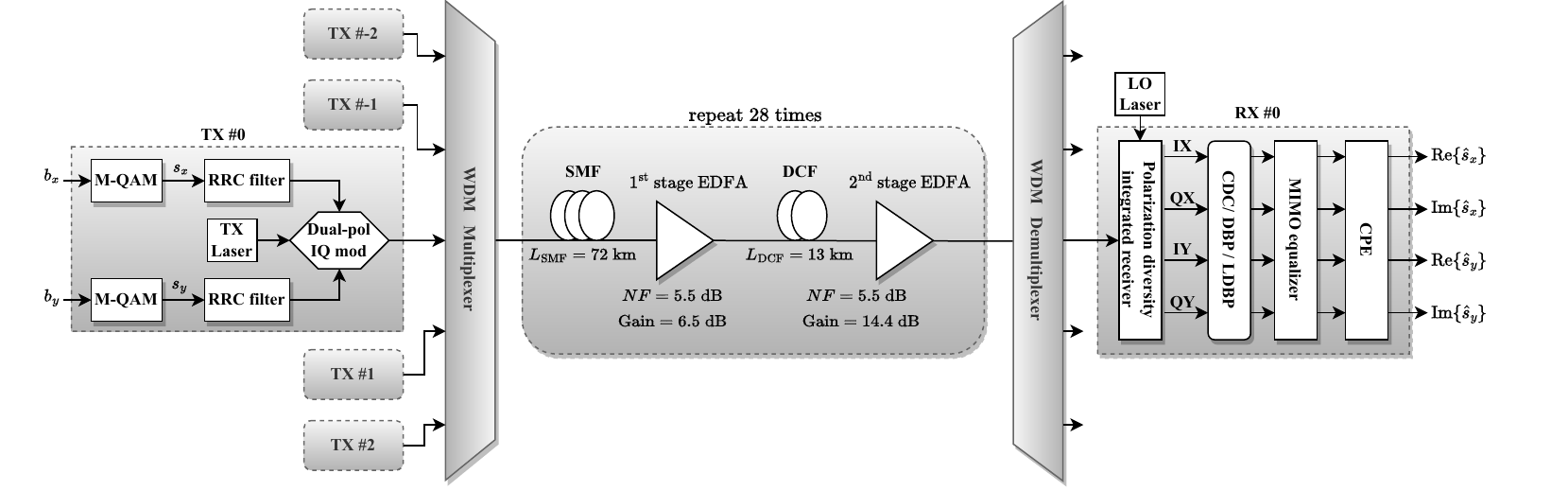}
    \caption{The DM-WDM optical fiber system.}
    \label{fig:sysmodel}
\end{figure}

\subsection{Transmitter and Channel}
In the polarization-multiplexed fiber transmission, two random bit streams with $N_b$ bits, $\mathbf{b_x}=(b_x^{(1)},b_x^{(2)},\dots,b_x^{(N_b)})$ and $\mathbf{b_y}=(b_y^{(1)},b_y^{(2)},\dots,b_y^{(N_b)})$, $b_{x/y}^{(i)}\in\{0,1\}$, are generated at the TX, then each is mapped into a sequence of $N_s$ symbols $\mathbf{s_x} = (s_x^{(1)},s_x^{(2)},\dots,s_x^{(N_s)})$ 
and $\mathbf{s_y} = (s_y^{(1)},s_y^{(2)},\dots,s_y^{(N_s)})$, where $s_{x/y}^{(i)}$ are drawn from 
the  constellation $\mathcal{S}$.
The Gray mapping is used to map the bit stream to the symbols in the constellation.
The baseband signal of the polarization $x$ is obtained by modulating $\mathbf{s_x}$ into the waveform
\begin{equation}
    q_{x,n}(z=0,t) = \sum_{i=1}^{N_s}s_{x,n}^{(i)}\,p(t-iT_s),
    \label{eq:single_channel}
\end{equation}
where $q_{x,n}$ is the complex envelope of the signal in the WDM channel $n$ as a function of distance $z$ 
and time $t$, $p(\cdot)$ is the pulse shape, and $T_s$ is the symbol period. 
$z=0$ indicates that the signal is located at the TX.
The equations for $q_{y,n}(0,t)$ are identical upon replacing $x$ with $y$. 
The waveforms $q_{x,n}$ and $q_{y,n}$ are multiplexed into an optical signal $\mathbf q_n(0,t)=[q_{x,n},q_{y,n}]$ using a dual-polarization Mach-Zehnder modulator.
The complex envelope of the WDM signal $\mathbf q(0,t)$ launched in the fiber link is generated  
by adding the signals of the different WDM channels 
\begin{equation}
    \mathbf q(0,t) = \sum_n \mathbf q_n(0, t) e^{-jn\Delta \omega t},
\end{equation}
where $\Delta\omega$ is the frequency spacing between adjacent WDM channels. 
The WDM signal is launched through an optical fiber channel consisting of multiple spans, where each span includes a SMF, dual-stage EDFAs, separated by a DCF with a proper length to compensate for dispersion. DCF has a higher nonlinearity compared to SMF, so the first-stage DCF provides pre-DCF gain, but only to a power level that would not generate excessive nonlinear effects. The second EDFA in the span then amplifies the signal further to its original power level.
The end-to-end channel with all components can be described by the interplay between CD, PMD and Kerr nonlinearity effects.
The propagation of signals inside each span in the presence of PMD is governed by the coupled nonlinear Schr\"odinger's equations (CNLSE), modeling the interaction between the two states of polarization

\begin{subequations}
\label{eq:CNLSE}
\begin{align}
\frac{\partial q_x}{\partial z} &= \big[-\frac{\alpha}{2}-\beta_{1x}\frac{\partial}{\partial t} - \frac{j\beta_2}{2}\frac{\partial^2}{\partial t^2} + j\gamma (|q_x|^2+\frac{2}{3}|q_y|^2) \big] q_x, \label{eq:cnlseX} \\
\frac{\partial q_y}{\partial z} &= \big[-\frac{\alpha}{2}-\beta_{1y}\frac{\partial}{\partial t} - \frac{j\beta_2}{2}\frac{\partial^2}{\partial t^2} + j\gamma (|q_y|^2+\frac{2}{3}|q_x|^2) \big] q_y, \label{eq:cnlseY}
\end{align}
\end{subequations}
where $\alpha$ is the loss parameter, $\beta_{1x/y}$ are the first-order dispersion coefficients, $\beta_{2}$ is the CD coefficient and $\gamma$ is the Kerr nonlinearity parameter. 
We consider a highly birefringent fiber, where the fiber length $\mathcal L$ is much larger than the  beat length $L_B={2\pi}/(\beta_{0x}-\beta_{0y})$, and the coherent cross-polarization terms can be neglected \cite[Eq.~6.1.11--12]{agrawal2021fiber}.
The effect of the rapidly changing state of polarization (SOP) is described separately in the numerical simulation of the propagation equation in Section \ref{sec:SSFM}.
At the end of the span, an EDFA with gain $G$ compensates for the fiber attenuation and introduces ASE noise. The noise $n(t)$ is a band-limited white circularly-symmetric complex Gaussian process with the auto-correlation function
$E\bigl(n(t)n^*(t)\bigr) = \sigma_0^2 \delta_B(t-t')$, where $\delta_B(x)=B {\rm sinc}(Bt)$, 
$\sigma_0^2= \frac{1}{2} (G-1) B h\nu_0 {\rm NF} $, where $\nu_0$ is the carrier frequency, $B$ is the signal bandwidth, $h$ is Planck constant, and NF is the amplifier's noise figure.

\subsection{Split-Step Fourier Method}
\label{sec:SSFM}
In this section, we present a numerical approach for solving the CNLSE using the SSFM.
All fiber effects occur simultaneously and accumulate along the length of the fiber.
In SSFM, a fiber of length $\mathcal L$ is divided into $N_{seg}$ segments of short length $\delta_s=\mathcal L/N_{seg}$, which we refer to as the \emph{PMD correlation length}, for which the SOP at the end of the segment is uncorrelated with its initial state.
The channel effects can be assumed to occur individually and consecutively in each segment. 
Each step of the SSFM involves evaluation of three sub-steps: a linear step, a PMD step and a nonlinear step. Below, we describe the steps in the SSFM. 
\begin{enumerate}
\item \emph{Linear step}: 
    Solves for the signal loss and CD in the frequency domain. 
    Considering only the terms which include $\alpha$ and $\beta_2$ in Eq.~\eqref{eq:CNLSE}, we obtain
    \begin{equation}
        \hat q_{x/y}(z,\omega) \rightarrow \exp \left(-\frac{\alpha}{2}\delta_s +  \frac{j\beta_2}{2} \omega^2 \delta_s\right) \hat q_{x/y}(z,\omega),
    \end{equation}
    where $\hat q_{x/y}$ denotes the Fourier transform of the time domain signal $q_{x/y}$.

\item \emph{PMD step}: 
The PMD can be modeled by applying the unitary Jones matrix $\mathbf J^{(i)}(\omega)$ to 
the signal vector $\mathbf{\hat q}(z,\omega)=[\hat q_x(z,\omega),\hat q_y(z,\omega)]^\top$
    \begin{equation}
        \mathbf{\hat q}(z,\omega) \rightarrow \mathbf J^{(i)}(\omega) \mathbf{\hat q}(z,\omega),
        \label{eq:PMD}
    \end{equation}
    where $\mathbf J^{(i)}(\omega) = \mathbf R^{(i)}\mathbf D^{(i)}(\omega)$.
The $\mathbf R^{(i)}$ here is a unitary matrix of the form
    \begin{equation}
        \mathbf R^{(i)} = \begin{bmatrix}
        e^{j \frac{\phi_i}{2}} \cos{\theta_i} & e^{-j \frac{\phi_i}{2}}  \sin{\theta_i} \\
        -e^{j \frac{\phi_i}{2}}  \sin{\theta_i} & e^{-j \frac{\phi_i}{2}} \cos{\theta_i}
        \end{bmatrix},
    \end{equation}
where $\phi_i$ and $\theta_i$, $i\in\{1,2,...,N_{seg}\}$, are sequences of independent identically distributed (i.i.d.) random variables drawn from a uniform distribution from $[0,2\pi)$. Furthermore, $\mathbf D^{(i)}(\omega)$ is the differential group delay (DGD) matrix
    \begin{equation}
        \mathbf D^{(i)}(\omega) = \begin{bmatrix}
        e^{-j\omega\frac{\tau_i}{2}} & 0 \\
        0 & e^{j\omega\frac{\tau_i}{2}}
        \end{bmatrix},
    \end{equation}
    where the DGD parameters $(\tau_i)_{i=1}^{N_{seg}}$
    are taken to be as i.i.d. random variables drawn from the natural probability distribution $\mathcal N(0,\tau\sqrt{\delta_s})$, where $\tau$ is the characteristic constant of the channel called the \emph{PMD coefficient}.

\item \emph{Nonlinear step}: Solves for the signal nonlinear effects  by only considering the terms which include
    $\gamma$ in Eq.~\eqref{eq:CNLSE}. For the $x$ polarization, 
    \begin{equation} 
        q_x(z,t) \rightarrow\exp\left(j\gamma\delta_s\big(|q_x|^2 + \frac{2}{3}|q_y|^2\big)\right)q_x(z,t).
    \end{equation}
\end{enumerate}

In our simulations, we consider a symmetric SSFM, where the nonlinear step is applied in the middle of the two linear half-steps.

\begin{figure}
    \centering
    \includegraphics{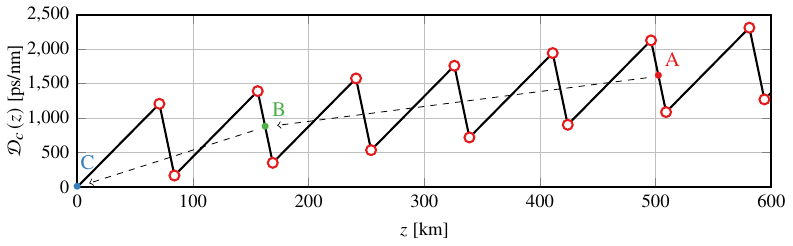}
    \caption{Dispersion map of the propagating optical signal over the first 7 spans, with two DBP steps shown. One DBP step is applied over 4 spans, while another is applied over 2 spans, with nonlinear operators applied at points $A$ and $B$. Accumulated dispersion is fully compensated at point $C$.}
    \label{fig:disp_map_and_power_dist}
\end{figure}

\subsection{Receiver}

At the RX, a polarization-diversity coherent receiver converts the optical signal to the electrical domain. 
A low-pass filter with the same bandwidth of the central WDM channel is applied to the signal, obtaining $\mathbf q_{0}(z=\mathcal{L},t)$. 
The resulting signal is sampled at 2 samples/symbol, and translated into four electrical signals corresponding to the $I$ and $Q$ components of each polarization, which are then processed by the DSP chain \cite{DSP_savory}.
The DSP chain consists of the following components.

\begin{enumerate}
    \item \emph{CD compensator (CDC)}, which reverses the CD as
    \begin{equation}
    \hat q_x(\mathcal L,\omega) \rightarrow \exp \Bigl(-j\frac{\bar\beta_2}{2}\omega^2\mathcal L\Bigr)\hat q_x(\mathcal L,\omega),
    \label{eq:CDC}
    \end{equation}
where $\bar\beta_2$ is the average group velocity dispersion in the link. According to Eq.~\eqref{eq:CDC}, CDC requires processing the signal in frequency domain using fast Fourier transform (FFT), followed by inverse FFT (IFFT) to transform the signal back to the time domain. An alternative implementation of the CDC in the time domain can be realized using FIR filters, which we will discuss in Section~\ref{sec:complexity}.

\item \emph{MIMO equalizer}, which compensates the time-varying PMD and the random SOP in the channel. 
    The adaptive equalization of both effects requires using a set of four real-valued finite impulse response (FIR) 
    filters, which together preform the inverse of the Jones matrix of the dynamic channel in Eq.~\eqref{eq:PMD}. The outputs of the MIMO equalizer are given by
    \begin{subequations}
    \begin{align}
        x_{\text{MIMO,out}}[k] = \mathbf h_{xx}^H\,\mathbf x_\text{in}[k]\, + \mathbf h_{xy}^H\,\mathbf y_\text{in}[k]\, \\
        y_{\text{MIMO,out}}[k] = \mathbf h_{yx}^H\,\mathbf x_\text{in}[k]\, + \mathbf h_{yy}^H\,\mathbf y_\text{in}[k]\,,
    \end{align}
    \end{subequations}
    where $\mathbf{h}_{xx}$, $\mathbf{h}_{xy}$, $\mathbf{h}_{yx}$, and $\mathbf{h}_{yy}$ are vectors of size $\xi$, representing the taps of the FIR filter, and $\mathbf{x}_{\text{in}}$ and $\mathbf{y}_{\text{in}}$ are sliding windows of the signal after chromatic dispersion (CD) compensation. These windows have a length of $\xi$, and they can be defined as $\mathbf{x}_{\text{in}}[k] = [x_\text{in}[k]$, $x_\text{in}[k-1]$, $\dots$, $x_\text{in}[k-\xi]]$ and $\mathbf{y}_{\text{in}}[k] = [y_\text{in}[k]$, $y_\text{in}[k-1]$, $\dots$, $y_\text{in}[k-\xi]]$.
    A popular method for optimizing the FIR taps is called the constant modulus algorithm (CMA), which is typically used for phase modulations. For the higher order QAM modulations we use in our simulations, the radially directed equalizer proposed in~\cite{DSP_savory} is used conjointly with the CMA. Following the MIMO equalizer, the output sequences are sampled at 1 sample/symbol.

\item \emph{Carrier phase estimation (CPE)}: The last step in the DSP is to estimate the PN, $\phi_N$, resulting from the phase fluctuations between the local oscillator at TX and RX ends \cite{Pfau:2009}. The PN of a single-frequency laser resembles a quasi-continuous frequency drift, and as result, the phase shift is characterized by a gradual and continuous change rather than discrete substantial phase jumps. A measure of the duration over which the laser phase remains stable is provided by the coherence time which is related inversely to the laser linewidth \cite{agrawal2021fiber}.
Within a single coherence time, the PN varies slowly compared to the signal, and can thus assumed to be constant. Therefore, 
    \begin{equation}
        x_{\text{sym}}[k] = x_{\text{CPE,in}}[k]\exp(-j\phi_N),
    \end{equation}
    where $x_{\text{CPE,in}}$ is the same as the output of the MIMO, $x_{\text{MIMO,out}}$.
\end{enumerate}

The aforementioned optical fiber receiver only concerns equalizing the linear effects of the signal resulting from the dispersive channel and PMD. 

\section{Digital Back-Propagation}

The signal propagation in the reverse direction through the optical fiber is numerically approximated using DBP. In Fig.~\ref{fig:sysmodel}, the DBP block takes the place of the CDC in the receiver (RX) and operates at the same sampling rate of 2 samples/symbol.
DBP employs the SSFM with negative propagation parameters and larger spatial segments compared to the SSFM. This approach helps to mitigate the high complexity typically associated with a large number of segments in the SSFM, while maintaining accurate signal reconstruction. In practice DBP is limited to 3 or less StpS.

\subsection{Mathematical Model}
The DBP block is placed before the MIMO equalizer, which means that the input signals are subject to PMD variations in the optical fiber. However, it is important to note that the standard DBP algorithm is not designed to equalize random effects like PMD. Instead, it primarily addresses deterministic effects such as CD and nonlinear effects.
moreover, the changing SOP can cause the $x$-polarization and $y$-polarization signals to be indistinguishable due to rapid changes in the orientation of the axes of birefringence $\theta$ along the fiber. When the PMD is small but the birefringence rapidly and varies randomly, the Manakov equation can be used to describe signal propagation. The propagation model we consider for DBP is based on Eq.~\eqref{eq:CNLSE} averaged over rapidly-varying SOP along the fiber, which corresponds to the vector Manakov equation\cite{Marcuse:1997}:
\begin{equation}
\frac{\partial \mathbf{q}}{\partial z} 
= - \frac{j\beta_2}{2}\frac{\partial^2  \mathbf{q}}{\partial t^2}
+
\Bigl[ 
-\frac{\alpha}{2}+
j\gamma \frac{8}{9}||\mathbf{q}||^2 \Bigr]\mathbf{q},
\label{eq:Manakov}
\end{equation}
Here, $\mathbf{q}(z,t) = [q_x(z,t), q_y(z,t)]^T$ is the Jones vector containing the propagating signals in both polarizations, and the factor $8/9$ is a characteristic property of the PMD, introduced by the averaging process.
In DBP, the optical channel is divided into $N_d$ spatial segments of equal lengths, denoted as $\delta_d = \mathcal{L}/N_d$. Within each segment, the dispersive and nonlinear channel effects are assumed to act independently. These effects correspond to the first and second terms on the right side of Eq.~\ref{eq:Manakov}, representing the linear and nonlinear effects, respectively. To solve for each of these terms, we use negated parameters. The linear part is solved in the frequency domain, while the nonlinear part is solved in the time domain, resulting in two partial solutions:
\begin{align}
\mathbf{\hat q}(z+\delta_d,\omega) &= \underbrace{\exp\left(-j\beta_2 \frac{1}{2} \omega^2 \delta_d \right)}_{L(\delta_d,\omega)} \mathbf{\hat q}(z,\omega), \\
\mathbf{q}(z+\delta_d,t) &= \underbrace{\exp\left( \frac{1}{2}\alpha \delta_d - j\frac{8}{9}\gamma ||\mathbf{q}(z,t)||^2 \delta_\text{eff}\right)}_{N(\delta_d,t)} \mathbf q(z,t).
\end{align}
where $\delta_\text{eff} = (1-\exp\{-\alpha \delta_d\})/\alpha$ represents the effective nonlinear step length, and the location $z$ is measured relative to the receiver, with $z=0$ corresponding to the receiver's location.
The DBP is hence characterized by two sets of operators: the linear operator $L(\delta_d,\omega)$ and the nonlinear operator $N(\delta_d,t)$. The equalized signal can be obtained from the received signal by alternating between these two solutions along the length of the fiber in the backward direction (from the receiver to the transmitter).

We assume a discretization of $\mathbf{q}(z,t) $ into the time-sampled vector $\mathbf{U}^{(n)} = [\mathbf X^{(n)},\mathbf Y^{(n)}]^T$, where $\mathbf X^{(n)}\in\mathbb C^N$ and $\mathbf Y^{(n)}\in\mathbb{C}^N$, respectively. The superscript $(n)$ refers to the DBP step, with $n=0$ representing the input to the DBP, and $n=N_d$ representing the output of the DBP. The linear step is represented by a matrix multiplication, given by
\begin{equation}
\mathbf{U}^{(n)} \rightarrow \mathbf{B}\mathbf U^{(n-1)} = \mathbf{W}^{-1}\text{diag}(e^{\delta_dH_1},...,e^{\delta_dH_n})\mathbf{W}\mathbf U^{(n-1)},
\label{eq:linear_param}
\end{equation}
where $\mathbf B \in \mathbb C^{N\times N}$ is a matrix representing the linear operator, $\mathbf{W}$ denotes the discrete Fourier transform matrix, $H_k = - j\beta_2 \omega_k^2/2$, and $\omega_k=2\pi f_k$, where $f_k$ corresponds to the $k$-th discrete frequency. Additionally, the nonlinear step can be represented by the nonlinear transformation $K(\cdot)$
\begin{equation}
    \mathbf{U}^{(n)} \rightarrow K\left(\mathbf U^{(n-1)}\right) = \mathbf U^{(n-1)} \exp\left(-j\gamma \varepsilon \delta_\text{eff}\frac{8}{9}\Bigl(\mathbf X^{(n-1)} \odot \mathbf{X}^{(n-1)*}+\mathbf
    Y^{(n-1)}\odot\mathbf{Y}^{(n-1)*}\Bigr)\right).
    \label{eq:nonlinear_param}
\end{equation}
where $\odot$ denotes the Hadamard product. Additionally, a real-valued parameter $\varepsilon \in [0,1]$ is introduced, to accurately model the nonlinear effects. The optimization of this parameter will be discussed in the results section.

\subsection{DBP Adaptation to DM Systems}

In NDM systems, CD is introduced by a single type of fiber, and the residual CD grows at a constant rate along the link. However, in DM systems, the compensation of CD using DBP becomes more involved as the linear step in DBP must account for the total dispersion generated by both the DCF and the standard SMF. The DBP algorithm we use in our work allows for flexible selection of StpS values less than 1, enabling us to include multiple spans in a single DBP step, similar to what is considered in \cite{647ecab0e,Gao:2021,Napoli:2014}. To model the accumulated dispersion in the fiber, we denote by $\mathcal{D}_c(z)$ the total accumulated dispersion inside the fiber as a function of distance $z$; see Fig.~\ref{fig:disp_map_and_power_dist}.
The linear step is adjusted as follows: Let us assume $N_d+1$ spatial steps, dividing a fiber of length $\mathcal{L}$ into spatially equal segments with step size $\delta_d = \mathcal{L}/N_d$, with the exception of the first and last steps where each has length $\delta_d/2$. Each step spans $[z_k,z_{k+1}]$, where $z_k=(k-\frac{1}{2})\delta_d$, $k \in \{1,\dots,N_d\}$, $z_0=0$, $z_{N_d+1}= \mathcal L$. This configuration is similar to the Wiener-Hammerstein model in \cite{millar_DBP}.
Within each step, we calculate the weighted-average dispersion coefficient, which is described by the following equation
\begin{equation}
\bar D = \frac{\mathcal D_c(z_k) - \mathcal D_c(z_{k-1})}{\delta_d}.
\label{eff_disp_coeff}
\end{equation}

The equation Eq.~\eqref{eff_disp_coeff} is essentially approximating the dispersion map between points $z_k$ and $z_{k+1}$ with a linear dispersion map with a $\bar D$ that is between the values of $D$ for SMF and DCF.
The power injected at the input of the DCF is set small enough to guarantee a quasi-linear transmission regime. As a consequence, the nonlinear step is performed with the coefficient $\bar\gamma=\gamma_{\text{SMF}}$ determined by the SMF.
This approximation is accurate, as shown in the numerical simulations that will be presented in Sec.~\ref{sec:results} of the paper.
The linear and nonlinear steps in the proposed DBP alternate until the algorithm spans over the entire optical link.

\subsection{Time domain and Frequency Domain Implementation of DBP}

The conventional method for implementing DBP involves using FFT and IFFT for each step, which can be computationally demanding due to the numerous FFTs and IFFTs required. However, considering the relatively low accumulated dispersion at the receiver in DM systems, we are interested in investigating whether a time-domain implementation of DBP could provide a complexity advantage over the frequency-domain approach. In this time-domain implementation, we replace the parameter $\mathbf{B}$ in Eq.~\ref{eq:linear_param} by employing an FIR filter with complex-valued taps, denoted as $h_\text{CDC}(\delta_d)$. This filter performs circular convolution with the backpropagating signal to compensate for the dispersion introduced within a step of length $\delta_d$. The DBP step in this case can be represented as \cite{Fougstedt:2017}
\begin{equation}
    \mathbf q(z+\delta_d,t) = (\mathbf q(z,t)*h_\text{CDC}(\delta_d))\cdot \exp(\alpha\delta_d/2)\cdot\exp(-j\delta_\text{eff}\varepsilon\gamma||\mathbf q||^2),
    \label{eq:TD-DBP}
\end{equation}
where $h_\text{CDC} = (h_{-F},\dots,h_{-1},h_0,h_1,\dots,h_F)$, and each $h_i$ for $i=1,2,\dots,F$ denotes an individual tap. It is worth noting that the filter taps exhibit symmetry, such that $h_i=h_{-i}$. The minimum number of taps needed in the FIR filter to compensate for dispersion within a DBP step is mainly determined by the length of the impulse response, approximated using the formula provided in \cite{Spinnler:2010}
\begin{equation}
    \tau_{CD}(\delta_d) = \frac{\lambda_c^2}{c}|D_{acc}|\Delta f,
\end{equation} 
where $|D_{acc}| = \bar D\,\delta_d$ is the total accumulated dispersion inside a step with length $\delta_d$, $\Delta f$ is the signal spectral width for a single channel, $\lambda_c=c/f_c$ is the carrier wavelength, and $c$ is the speed of light. The channel impulse response length $\tau_{CD}(\delta_d)$ is measured in seconds, in which case, the number of filter taps for any step is
\begin{equation}
    N_{CDC,\delta_d} = \left\lceil \frac{\tau_{CD}(\delta_d)}{T_{s}}n_s\right\rceil,
    \label{eq:number_of_taps}
\end{equation}
and $\lceil x\rceil$ denotes the smallest integer larger or equal to $x$, and $n_s$ is the oversampling ratio.

\section{Learned Digital Back-Propagation}

\begin{figure}
    \centering
    \includegraphics[width=\linewidth,trim={1.6cm 0 2.7cm 0},clip]{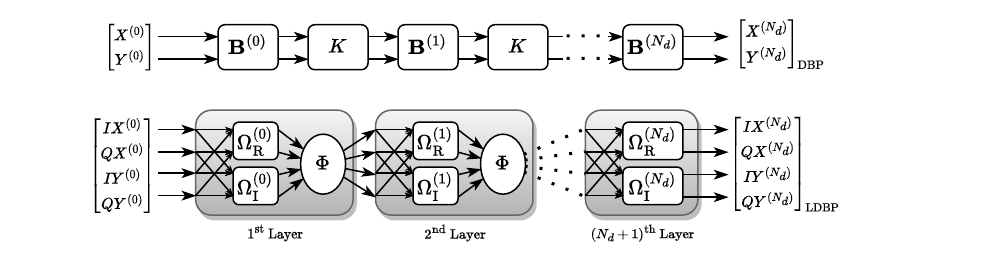}
    \caption{Block diagram of DBP (upper branch) and LDBP (lower branch) structures. The following symbols represent: $\mathrm{IX}\coloneqq\text{Re}\{\mathbf X\}$, $\mathrm{QX}\coloneqq\text{Im}\{\mathbf X\}$, $\mathrm{IY}\coloneqq\text{Re}\{\mathbf Y\}$, and $\mathrm{QY}\coloneqq\text{Im}\{\mathbf Y\}$.}
    \label{fig:LDBP-block-diagram}
\end{figure}

When designing NNs, incorporating prior knowledge of the system's model can significantly expedite the training process and improve the convergence rate to a lower local minimum compared to black-box approaches \cite{sergei_CNN}.
The similarity between the SSFM and deep feed-forward NNs had been pointed out in the literature \cite{Hagers_first,Hager_butler_LDBP}, where both involve alternating between linear matrix multiplication and nonlinear element-wise operator. This similarity can be exploited to design a model-based NN with DBP as a blueprint, which allows for optimizing DBP parameters using the SGD algorithm \cite{Hagers_first}.

\subsection{Mathematical Model of NNs}
Deep feed-forward NNs consist of an input layer, output layer, and a cascade of hidden layers each performing a nonlinear transformation on the input of the layer, then passing the output to the next layer. 
Deep feed-forward NNs can be mathematically represented as a series of alternating linear operations, denoted as $\mathbf{A}^{(k)}$, and element-wise activation function, denoted as $\Phi^{(k)}$. These operations create a mapping from an input vector $\mathbf{u}$ to an output vector $\mathbf{v}$, as follows:
\begin{equation}
\mathbf{v} = \Phi^{(N_l)}(\mathbf{A}^{(N_l)}(\Phi^{(N_l-1)}(\dots \mathbf{A}^{(0)}(\mathbf{u})))),
\end{equation}
where $N_l$ represents the number of layers in the NN, and the superscript $(k)$ denotes the index of the $k$-th layer of the NN. The function performed by the linear operator $\mathbf{A}^{(k)}$ varies depending on the neural network architecture. In the case of fully-connected neural networks (NNs), it performs matrix multiplication. However, in the case of convolutional NNs, it performs convolution.

In our application for signal equalization, we specifically utilize the convolutional NN model. Within this model, the linear operator $\mathbf{A}^{(k)}(\mathbf{c})$ in each layer is defined as $\mathbf{A}^{(k)}(\mathbf{c}) = \mathbf{c} \ast \mathbf{\Omega}^{(k)} + \mathbf{b}^{(k)}$, where $\mathbf{\Omega}^{(k)}\in\mathbb{C}^{m}$ represents the convolutional filter, and $\mathbf{b}^{(k)}\in\mathbb{C}^{N}$ is the bias vector.

\subsection{Architecture of LDBP}

Fig.~\ref{fig:LDBP-block-diagram} depicts the block diagrams of LDBP and its blueprint DBP. The NN architecture of LDBP is a complex-valued network that comprises two real-valued networks operating jointly. Each of these networks contains $N_l=N_d$ layers and accepts four input vectors ($IX$, $QX$, $IY$, and $QY$). Each layer of the network consists of two parallel convolutional filters, $\Omega_R$ and $\Omega_I$, corresponding respectively to the real and imaginary parts of the filter, and a nonlinear function $\Phi$, which takes four input vectors and generates four output vectors. The number of non-zero weights in the convolutional filter is determined numerically, as will be described in the simulation setup section later. These layers perform the real-valued equivalent of the operation described in Eq.~\eqref{eq:TD-DBP} in the time domain without the need for FFT and IFFT. Biases are not utilized in our model and are set to zero. In our model, we do not use the biases and therefore set them to zero. The input dimension of LDBP is $[N_{ex},N,4]$, and its output dimension is $[N_{ex},N-2M,4]$, where $M$ represents the memory of the dispersive channel.

To train the Learned-DBP, we simulated a five-channel WDM PMD-free transmission of a block of $2^{15}$ symbols at various launch powers. The signal was initially sampled at 16 samples per symbol duration for forward propagation using SSFM. However, after the WDM demultiplexer, the signal from the central channel was sampled at a sampling ratio of $n_s = 2$ samples per symbol, resulting in a received block consisting of $2^{16}$ samples. The NN operates in a sliding window fashion, with a window size of $N$ sliding over the transmission block and advancing by $L\times n_s$, where the sliding factor $L$ is an integer that determines the number of symbols shifted between examples.
For LDBP training, we generated input-output pairs with a size of 1024 samples (512 symbols) for the input, and the corresponding output was 852 samples (426 symbols) long due to the dispersive channel effects (43 symbols on each side). We set the sliding factor to $L=8$, generating $N_{ex}=4021$ input-output pairs for training.
To test LDBP, we generated 8 transmission blocks with PMD using the same overlapping and shifting technique as for training. However, this time, the sliding factor was set to $L = 426$ to ensure that each symbol in the transmission block was detected exactly once. The output of LDBP was then passed to the DSP to equalize PMD and dynamic channel effects.

We implement a symmetric DBP as the blueprint of LDBP, such that all layers are initialized with the corresponding parameters in a linear step of DBP at the full step size $\delta_d$, except for the first and last layers which correspond to a half  step $\delta_d/2$. 
The NN is trained by minimizing the mean squared error (MSE) loss function, using the \emph{Adam} optimizer with a learning rate of 0.001. During training, 20\% of the training examples were used for validation to monitor the LDBP's progress. The LDBP was trained for up to 75 epochs, with an early-stop condition triggered if the validation error did not decrease within 5 epochs. The best-performing epoch's weights were used in the final LDBP. After training at each launch power, we evaluate the performance of the LDBP by calculating the Q-factor using independently generated testing data.

\begin{table}[t]
    \centering  
    \small
    \begin{tabular}{cccc}
        \hline
        & \# of channels & Modulation format & fiber coefficients \\
        \hline
        \hline
        Setup (A) & 1  & $16$-QAM & \shortstack{$\alpha$=0.2 dB/km, PMD = 0.05 ps/$\sqrt{\text{km}}$,\\$D$=17 ps/(nm.km), $\gamma$= 1.4/W/km} \\
        Setup (B) & 5 & $16$-QAM & \shortstack{\phantom{text}\\Same as above}\\
        Setup (C) & 5 & $64$-QAM & \shortstack{\phantom{text}\\Same as above}\\
        Setup (D) & 5 & $16$-QAM &  \shortstack{$\alpha$=0.24 dB/km, PMD = 0.3 ps/$\sqrt{\text{km}}$,\\$D$=17 ps/(nm.km), $\gamma$= 1.4/W/km}\\
        \hline
    \end{tabular}
    \caption{Description of the simulated setups.}
    \label{setups}
\end{table}

\section{Simulated System Setup and Performance Results}

\label{sec:results}

The performance results are based on the simulation of the transmission system shown in Fig.~\ref{fig:sysmodel}.
All elements of the transmission system, including the transmitter (TX), receiver (RX), and channel, are simulated in Python, while the NN is implemented using the TensorFlow library.
In this section, we present the performance results of DBP and LDBP for four different setups: (A), (B), (C), and (D). Setup (A) represents a single-channel transmission system with 16-QAM modulation, while setups (B) and (C) are WDM transmission systems with 16-QAM and 64-QAM modulation formats, respectively. Setup (D) is also a WDM transmission using 16-QAM modulation, but it includes aging effects where a fiber channel undergoes aging. The aging study will be described in detail when we present the performance results for this setup. Table~\ref{setups} provides a detailed description of each setup.

For all setups (A)--(D), the transmission symbol baud rate $B=32$~GBaud for each channel using a root-raised-cosine (RRC) pulse-shape with a roll-off factor $\rho=0.06$. The optical fiber link consists of $N_{sp}=$~$28$ spans, each span including an SMF and a DCF measuring $72$~km and $13$~km, respectively.
The length of DCF is chosen such that it compensates for $85$\% of the CD in each span.
An amplifier with gain $G_{\text{SMF}} = 6.5$ dB is applied at the end of the SMF, and a second amplifier with gain of $G_{\text{DCF}} = 14.4$ dB is applied after the DCF. 
The SMF parameters and PMD value for all setups are provided in Table~\ref{setups}.
For WDM setups (B)--(D), the WDM channels are separated by a frequency spacing of $37.5$~GHz, resulting in a guard band of 5.5 GHz between adjacent channels.
The lasers used for these setups had a linewidth of 50 kHz. To avoid overestimation of nonlinear crosstalk, the data symbols of all WDM channels were intentionally made unsynchronized in terms of time, polarization state, and phase. 
At the receiver, an RRC filter with a bandwidth of $(1+\rho)B$ is applied to filter out the adjacent channels, such that only the central channel is processed by the DBP or LDBP algorithms.
The forward signal propagation using SSFM is simulated with 72 steps for each SMF and 13 steps for each DCF. The signals are sampled at a rate of 16 samples per symbol. At the receiver, the signal is downsampled to twice the symbol rate before being processed by either the CDC, DBP, or LDBP algorithms. Finally, the output of the CDC, DBP, or LDBP is downsampled to one sample per symbol and processed by the conventional DSP chain, which equalizes PMD effects and polarization mixing. 

\subsection{DBP Parameters Optimization}
\begin{figure}[t]
\begin{subfigure}{0.5\textwidth}
  \centering
  \includegraphics[width=\linewidth]{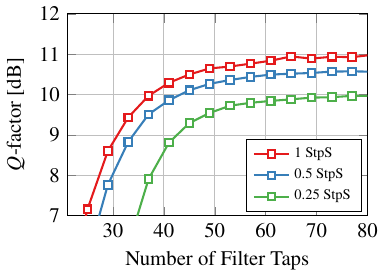}
  \caption*{(a)}
\end{subfigure} %
\begin{subfigure}{0.5\textwidth}
  \centering
  \includegraphics[width=\linewidth]{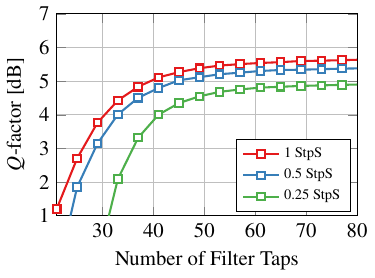}
 \caption*{(b)}
\end{subfigure}%
\caption{Achieved Q-factor by DBP at the optimal launch power (-2 dBm) for different values of StpS as a function of filter taps per step for (a) 16-QAM and (b) 64-QAM.}
\label{fig:impact_of_filter_taps}
\end{figure}
To optimize the performance of the DBP for all transmission scenarios (A)--(D), we select the parameter $\varepsilon$ in Eq.~\ref{eq:nonlinear_param} such that each DBP achieves the highest $Q$-factor at the optimal launch power. For a single channel transmission, we find that the value of $\varepsilon$ is 1 for all values of StpS.
For the WDM transmission of 5 channels (Set-ups B, C and D), we find that the optimal values are $\varepsilon = 0.85$ for DBP with 1 StpS, $\varepsilon = 0.75$ for DBP with 0.5 StpS, and $\varepsilon = 0.64$ for DBP with 0.25 StpS. It should be noted that $\varepsilon = 0$ corresponds to performing LE, and in this case, all DBP configurations perform similarly to LE, regardless of the number of StpS.
The filter size in each DBP step varies depending on the number of StpS, Optimal values of $F$ were determined for different DBP configurations by numerical optimization. Specifically, $F=16$ was found to be optimal for both 1 StpS and 0.5 StpS DBP, $F=24$ for 0.25 StpS, $F=30$ for $\frac{1}{7}$ StpS, and $F=36$ for $\frac{1}{14}$ StpS. These values represent the minimum values required to ensure that all DBP configurations outperform LE at all launch powers. The impact of filter width on performance is shown in Fig.\ref{fig:impact_of_filter_taps}. It is worth noting that this finding agrees with previous literature, specifically \cite{APTL_NN}, which reported that the filter width has a significant impact on the performance of DBP-based equalizers (see Fig.(9) in \cite{APTL_NN}).

\subsection{Performance Comparison of DBP and LDBP in Single Channel Transmission}
A single channel transmission is simulated in setup (A). We choose ($\text{SNR}_{\text{eff}}$) as a performance measurement for single channel transmission, which is defined as
\begin{equation}
\text{SNR}_{\text{eff}} = 
\frac{\rvert\rvert \mathbf{\hat s_x} \rvert\rvert^2+\rvert\rvert \mathbf{\hat s_y} \rvert\rvert^2}
{\rvert\rvert \mathbf{s_x} - \mathbf{\hat s_x} \rvert\rvert^2 + \rvert\rvert \mathbf{s_y}-\mathbf{\hat s_y} \rvert\rvert^2},
\end{equation}

\begin{figure}[t]
\centering
\begin{subfigure}{0.7\textwidth}
    \centering
    \includegraphics[width=\linewidth]{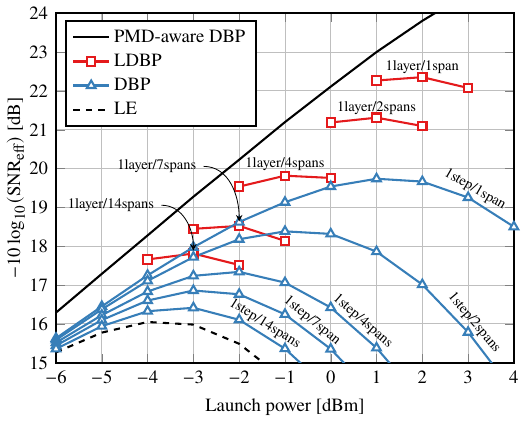}
\end{subfigure}
\caption{DBP performance for different StpS values in single channel transmission (setup A). The performance of LDBP is also shown for 3 launch powers around the peak $\text{SNR}_{\text{eff}}$.}
\label{fig:DBP_for_SC}
\end{figure}
where $\mathbf{\hat s_x}$ and $\mathbf{\hat s_y}$ denote the estimated symbols at the output of the DSP, with true values $\mathbf{s_x}$ and $\mathbf{s_y}$, respectively.
{The reason for choosing $\text{SNR}_\text{eff}$ is the high reliability of the single channel scenario, which leads to an extremely low bit-error ratio (BER). This makes it challenging to measure BER accurately, requiring a large number of training examples. The $\text{SNR}_\text{eff}$ provides a more practical measure of performance under these conditions.} 
The performance of the DM-adapted DBP and LDBP in setup (A) is depicted in Fig.~\ref{fig:DBP_for_SC} for varying signal launch powers. The PMD-aware DBP in Fig.~\ref{fig:DBP_for_SC} is a genie-aided model which assumes perfect knowledge of PMD and SOP across the fiber channel at the RX and uses them in the back-propagation with the same step size $\delta_s$ used for SSFM. Despite its impracticality, the PMD-aware DBP model provides a useful upper bound on the performance that can be achieved by PMD-agnostic DBP and LDBP algorithms. Our simulations consider DBP and LDBP equalizers with a fractional number of StpS or LpS that is less than or equal to 1.

At the optimal launch power, LDBP with 1 LpS outperforms DBP with the same complexity by providing an $\text{SNR}\text{eff}$ of 22.3 compared to 19.8 dB, with the optimal launch powers differing by about 1 dBm between the two algorithms. The LE achieves its best performance at an $\text{SNR}{\text{eff}}$ value of 16 dB, which is achieved with a launch power of -4 dBm. Both DBP and LDBP outperform the LE with varying gains, with LDBP with 1 LpS achieving the highest gain of 6.3 dB and DBP with 1 StpS achieving a gain of 3.8 dB. The simulated LDBP with the least complexity has 2 layers (1 full step and 2 half steps) and 2 activation functions, and outperforms DBP with similar complexity by 1.4 dB and the LE by 1.8 dB.

\subsection{Performance Comparison of DBP and LDBP in Multi-Channel Transmission}
\begin{figure}
\begin{subfigure}{0.46\textwidth}
  \centering
  \includegraphics[width=\linewidth]{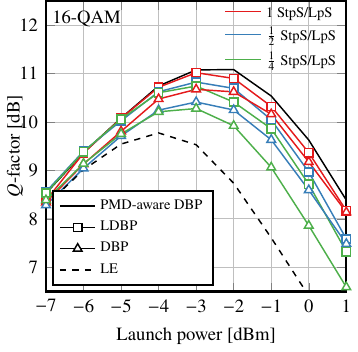}
 \caption{}
\end{subfigure} \hfill
\begin{subfigure}{0.46\textwidth}
  \centering
  \includegraphics[width=\linewidth]{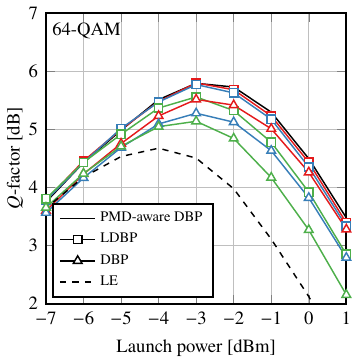}
 \caption{}
\end{subfigure}
\caption{Achieved $Q$-factors for DBP and LDBP with different values of StpS and LpS for WDM transmissions over a 2016 km fiber with 32 Gbaud using (a) 16-QAM modulation, and (b) 64-QAM modulation.}
\label{fig:main_result16QAM}
\end{figure}

Setups (B)--(D) present multiple WDM transmission scenarios. In such cases, the nonlinearity affecting the received signal is dominated by the nonlinear interference introduced by the adjacent channels via XPM. Since only the signal from one single channel is fed to the receiver, the information in adjacent channels is unknown to the receiver. Therefore, the nonlinearity generated by adjacent channels impacts all equalizers and limits their performance in the nonlinear regime. The performance of the PMD-aware DBP equalizer in these setups can be characterized by a bell-shaped curve, as seen in Figures \ref{fig:main_result16QAM} and \ref{fig:aged_fiber}(b), in contrast to the straight line observed in the single-channel scenario (A) shown in Fig.~\ref{fig:DBP_for_SC}.

For setups (B)--(D), we utilize the $Q$-factor based on BER as the performance metric for the implemented algorithms. The $Q$-factor is defined as follows
\begin{equation}
Q\textnormal{-factor}= 20\log_{10}[\sqrt{2}\text{erfc}^{-1}(2 \text{BER})],
\end{equation}
where $\text{erfc}(\cdot)$ is the complementary error function.

Simulation results for setups (B) and (C) are presented in Fig.~\ref{fig:main_result16QAM}. In these setups, the DBP and LDBP algorithms were simulated with varying numbers of total steps or layers ($M \in {7,14,28}$).

In setup (B), LDBP with 1 StpS achieved a peak $Q$-factor of 11 dB at the optimal launch power of $P=-3$ dBm. Notably, the peak performance of LDBP with 1 StpS is comparable to that achieved by the PMD-aware DBP at the same launch power. When comparing LDBP to DBP with 1 StpS, LDBP outperformed DBP by 0.3 dB at the same launch power. Furthermore, LDBP with 0.5 StpS and 0.25 StpS achieved $10.8$ dB and $10.7$ dB, respectively, both outperforming DBP with similar numbers of StpS by 0.4 dB.

In setup (C), The LDBP with 1 StpS achieved a peak $Q$-factor of 5.8 dB, which is 0.3 dB higher than the peak performance achieved by the DBP with a similar number of StpS. The LDBP with 0.5 StpS and 0.25 StpS achieved a peak $Q$-factor of 5.8 dB and 5.6 dB, respectively, both outperforming the DBP with a similar number of StpS by 0.5 dB. On the other hand, the LE achieved a peak $Q$-factor of 4.7 dB.

It is worth noting that the launch powers corresponding to the peak $Q$-factor in both Fig.\ref{fig:DBP_for_SC} and Fig.\ref{fig:main_result16QAM} are shifted by 4 dB for WDM transmission compared to single-channel transmission. This is due to the XPM generated by the four adjacent channels, which dominates the nonlinearity affecting the received signal in WDM setups, unlike single-channel transmission, which only experiences self-phase modulation (SPM) from the same channel, as explained in \cite{essiambre}.

\subsection{Impact of Aging Effects on LDBP Performance}
\begin{figure}[t]
    \begin{subfigure}{0.46\textwidth}
        \centering
        \includegraphics[width=0.94\linewidth]{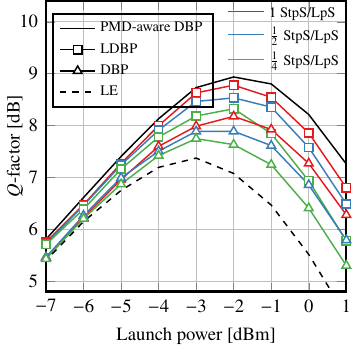}
        \caption{}
    \end{subfigure}
    \hfill
    \begin{subfigure}{0.46\textwidth}
        \centering
        \includegraphics[width=\linewidth]{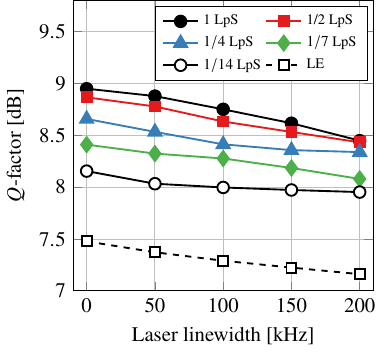}
        \caption{}
    \end{subfigure}
    \caption{Achieved $Q$-factors for LDBP and DBP with different StpS and LpS values for Set-up (D) simulating transmission over a 2016 km aged fiber at 32 Gbaud. Figure (a) shows $Q$-factors across launch powers, and (b) displays the impact of laser PN on the peak $Q$-factor of LDBP with varying LpS values.}
    \label{fig:aged_fiber}
\end{figure}
Optical fiber links deployed for intensity modulation and direct detection (IM-DD) communication, especially those utilizing dispersion management, are particularly susceptible to aging and poor maintenance due to being outdated systems \cite{1497628,Bohata:2017}. Changes in the fiber material occur as the fibers age, leading to the deterioration of their parameters over time. Additionally, yearly temperature variations can contribute to higher values of differential group delay (DGD). Additionally, fiber splicing, which is done during maintenance, can increase attenuation and necessitate higher amplification gain in the link, resulting in ASE noise. To reflect these changes in a more realistic transmission scenario, we consider the impact of aging and maintenance-induced changes on the fiber link.
It is worth noting that the CD coefficient $\beta_2$ and the nonlinearity coefficient $\gamma$ are not affected by aging effects. While only these two coefficients are used to initialize the LDBP, the new attenuation coefficient $\alpha$ in aged fibers can cause a reduction in the effective length $L_\text{eff}$. Thus, retraining the NN with the updated parameters is necessary. Our goal is to examine how these changes affect the LDBP's performance relative to DBP, and compare the two methods in a realistic transmission scenario that incorporates aging and maintenance-induced changes in the fiber link.

The fiber parameters selected for setup (D) aim to simulate aging effects and consist of an attenuation coefficient of $\alpha = 0.24$ dB/km and a PMD coefficient of $0.3$ ps/km$^{-1/2}$. The value of the attenuation coefficient after degradation is based on experimental findings reported in \cite{Bohata:2017}. The simulation results for this particular setup are presented in Figure \ref{fig:aged_fiber}.
To account for the changes in the fiber link, we retrained the model using data generated from simulations with the new parameters. The LDBP with 1 StpS achieved a peak $Q$-factor of $8.8$ dB at a launch power of $P=-2$ dBm, which is $0.6$ higher than the peak $Q$-factor of the DBP with similar complexity.
Furthermore, we compared setup (D) with setup (B), which simulates a similar transmission scenario without aging effects. We observed an average drop in $Q$-factor of 2.2 dB across all graphs, but the $Q$-factor gain of LDBP over DBP was still maintained. This suggests that deploying the LDBP long-term is feasible with retraining of the model.

The existing DM systems deployed for IM/DD may not be optimized for coherent transmission. In fact, lasers used in IM/DD systems typically have higher laser PN than those used in coherent transmission, which can potentially impact the performance of receivers used for coherent detection \cite{Fatadin:2010}. In general, laser phase noise does not have a significant impact on the system, as the time interval over which the laser phase noise changes is much longer than the symbol period.
However, to investigate its impact on the performance of LDBP for set-up (D), we choose to retrain the neural network using data that incorporates varying degrees of laser PN. While it is the primary role of the digital signal processing (DSP) following the LDBP to mitigate dynamic effects like laser PN, when the LDBP is exposed to examples of data affected by PN, it may be possible for the neural network to learn and reduce the impact of PN to some extent. 
The peak $Q$-factor achieved for different values of LpS and a range of laser linewidths is shown in Fig.~\ref{fig:aged_fiber}(b). Since the input vector for LDBP has a width of 512 symbols, the deviation in PN for the laser linewidth used in the simulations is approximately 0.01 radian, which is too small to significantly impact the performance. The results indicate that the average drop in $Q$-factor is 0.1 dB for a laser linewidth of 100 kHz and 0.2 dB for a laser linewidth of 200 kHz.

\section{Complexity of DBP and LDBP}
\label{sec:complexity}
We quantify the complexity of each step in terms of the number of real multiplications (RMs) per detected symbol (RMpS), excluding additions. It is important to note that both DBP and LDBP exhibit the same complexity when considering the same number of StpS and LpS. Therefore, the complexity formulas derived for LDBP in this section are identical to those for DBP.

To efficiently compute the exponential function in the activation function, various approximation algorithms can be employed, such as the CORDIC algorithm \cite{CORDIC1, CORDIC2}. These algorithms utilize look-up tables and bit-shifts, eliminating the need for multiplications. By employing similar algorithms, the activation function can be computed efficiently, with each computation of the activation requiring only 9 RMs.

We make a distinction between the complexities of TD-LDBP and FD-LDBP and present their respective complexity formulas as follows:
\begin{figure}[t]
\begin{subfigure}{.47\textwidth}
    \centering
    \begin{picture}(200,83)
    \put(0,0){\includegraphics[width=1\linewidth]{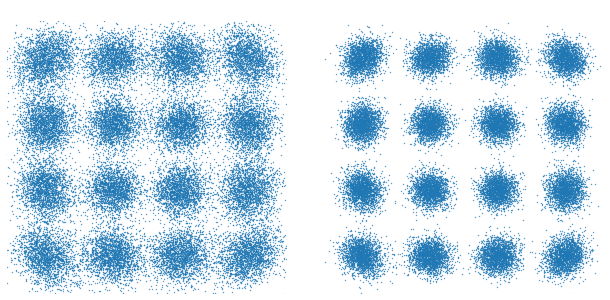}}
    \put(39,83){\footnotesize{LE}}
    \put(127,83){\footnotesize{LDBP}}
    \end{picture}
    \caption*{Set-up B}
\end{subfigure}
\hspace{1cm}
\begin{subfigure}{.47\textwidth}
    \centering
    \begin{picture}(200,83)
    \put(0,0){\includegraphics[width=\linewidth]{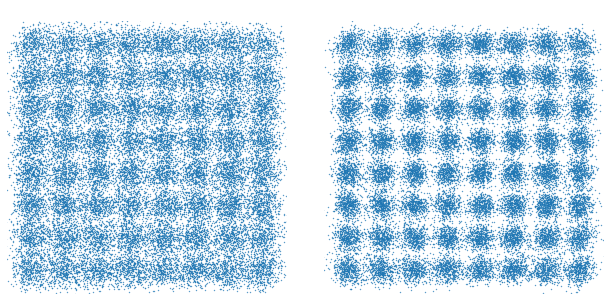}}
    \put(39,83){\footnotesize{LE}}
    \put(127,83){\footnotesize{LDBP}}
    \end{picture}
    \caption*{Set-up C}
\end{subfigure}
\caption{Constellation diagrams for LE and LDBP at their respective optimal launching powers for Set-ups B\&C. These constellations were obtained at the end of the DSP prior to hard-decision.}
\label{fig:constellation}
\end{figure}
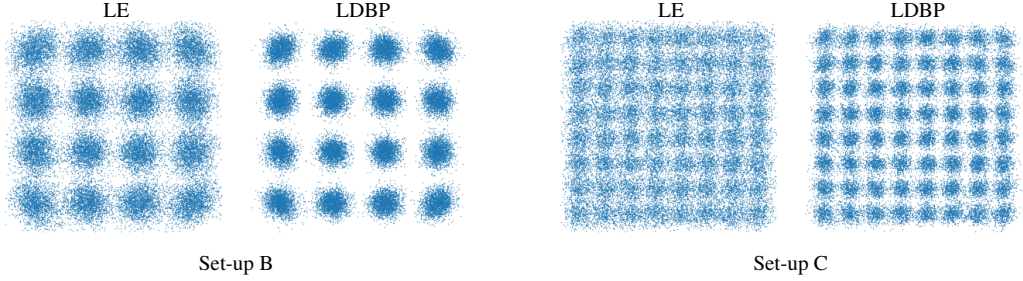

\paragraph{FD-LDBP Complexity} 
The total complexity of FD-LDBP per detected symbol, can be measured in RMpS for a signal with input size $N$ and sampled at rate $n_s$ samples/symbol. The complexity is given by~\cite{sergei_CNN}
    \begin{equation}
        C_{\text{FD--LDBP}} = \left(N_d+1\right)\left(4\frac{N(\log_2(N)+1)n_s}{N-N_{CDC,\mathcal{L}}+1}\right) + \frac{9}{2}N_dn_s.
    \end{equation}

\paragraph{TD-LDBP Complexity} The complexity of TD-LDBP for each detected symbol is determined by the convolution of a kernel with size $F$ and an input of size $N$, which involves $4FN$ RMs. This computation assumes a dilation of 1, stride of 1, and padding of $F-1$, resulting in an output size that is the same as the input size for each convolutional layer. Therefore, the total complexity per detected symbol for TD-LDBP can be calculated as
\begin{equation}
    C_{\text{TD--LDBP}} = \left(N_d+1\right)\left(4\frac{(2F+1)Nn_s}{N-N_{CDC,\mathcal{L}}+1}\right) + \frac{9}{2}N_dn_s.
\end{equation}
\begin{figure}[t]
    \centering
    \includegraphics[width=\linewidth]{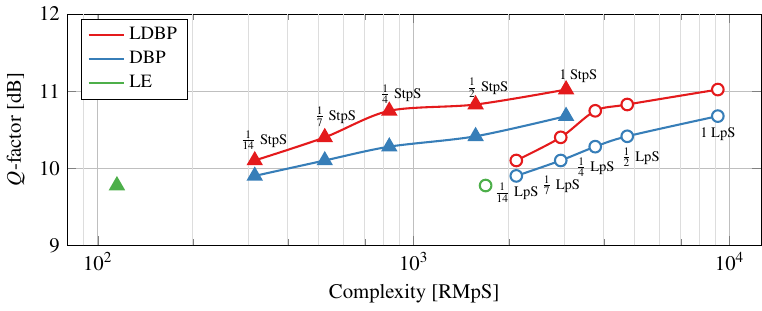}
\caption{Trade-off between complexity and performance for different equalizers in both FD-LDBP (represented by triangles) and TD-LDBP (represented by circles).}
\label{fig:Complexity}
\end{figure}

Fig.~\ref{fig:Complexity} depicts the RMpS complexity of DBP and LDBP using both TD and FD implementations. The complexity of FD-LDBP (or FD-DBP) is primarily dependent on the number of steps involved in the algorithm, and can be approximated by 114 times the number of FFT/IFFT uses. The complexity of 1 LpS FD-LDBP, which uses 29 FFT/IFFT pairs, is approximately 3300 RMpS. On the other hand, TD-LDBP with the same number of LpS has a complexity of around 9000 RMpS. In fact, FD-LDBP is less computationally complex compared to TD-LDBP across all values of LpS. We initially hypothesized that smaller finite impulse response (FIR) filters would be required to perform each linear step in DM compared to NDM systems, given that the accumulated dispersion in our DM system was 85\% lower than that of NDM. This led us to believe that TD-implementation might be less complex than FD-implementation. However, we discovered that strictly following Eq.~\ref{eq:number_of_taps} to compute the number of taps in each FIR filter did not work. The error resulting from truncation in each step accumulated with each consequent step in the DBP, leading to errors that caused DBP to perform worse than LE. We determined the number of filter taps required in each step through numerical simulations, as demonstrated by Fig.~\ref{fig:impact_of_filter_taps}. Our results showed that TD implementation did not offer any gain in terms of complexity compared to FD implementation. Nevertheless, TD implementation may offer a more hardware-friendly approach for field-programmable gate arrays (FPGAs) and similar devices that rely on multiply and accumulate (MAC) operations, as it does not require an FFT/IFFT operation.

\section{Conclusions}
In this paper, we have presented the LDBP approach for mitigating nonlinear effects in DM optical fiber transmission systems. LDBP is a DBP optimization method that leverages NNs training algorithms to optimize the DBP parameters. Our comparative study has shown that LDBP outperforms DBP, providing a significant gain in Q-factor, with an average improvement of 0.4 dB. The application of NNs to DM links is an important new development, as it demonstrates the possibility of repurposing DM systems for coherent transmission. The results have significant implications for the fiber-optics industry, suggesting that data rates in conventional DM optical links can be substantially improved using modern, simple coherent receivers.
Additionally, we demonstrated the complexity of both DBP and LDBP using two techniques, TD and FD. Contrary to our initial belief, TD-implementation was found to be more complex across all values of LpS. 
Overall, our findings highlight the potential of LDBP as an effective method for mitigating nonlinear effects in DM optical fiber transmission systems.

\begin{backmatter}
\bmsection{Funding}
Content in the funding section will be generated entirely from details submitted to Prism. Authors may add placeholder text in the manuscript to assess length, but any text added to this section in the manuscript will be replaced during production and will display official funder names along with any grant numbers provided. 

\bmsection{Acknowledgments}
This work has received funding from the EU Horizon 2020 program under the Marie Sk\l{}odowska-Curie grant agreements 813144, and the European Research Council (ERC) research and innovation programme, Grant Agreement No. 805195. 
A. Napoli, N. Costa and J. Pedro would like to thank the European Commission for funding their activities through the H2020 B5G-OPEN (G.A. 101016663). 

\bmsection{Disclosures}
The authors declare no conflicts of interest.

\bmsection{Data availability} Data underlying the results presented in this paper are not publicly available at this time but may be obtained from the authors upon reasonable request.

\end{backmatter}


\end{document}